\documentclass[preprint]{revtex4}

\usepackage{graphicx}
\newcommand{\eg}{{\it e.g.}}
\newcommand{\ie}{{\it i.e.}}

\begin{document}
\bibliographystyle{revtex}

\preprint{IIT-HEP-01/5}

\title{Introduction to Muon Cooling\footnote{Presented at the {\sl Snowmass Summer Study on the Future of Particle Physics}, Snowmass, Colorado, June 30--July 21, 2001.}
}



\author{Daniel M. Kaplan}
\email[]{kaplan@fnal.gov}
\affiliation{Illinois Institute of Technology, Chicago, IL 60616}

\author{for the Muon Collaboration}
\noaffiliation


\begin{abstract}
Starting from elementary concepts, muon-beam cooling is defined, and the techniques by which it can be accomplished introduced and briefly discussed.

\end{abstract}

\maketitle

\section{Motivation}

High-energy muon beams have been proposed as uniquely powerful and incisive sources for  neutrino scattering and oscillation studies.  They may also enable energy-frontier lepton-antilepton colliders and may have unique advantages for studying the physics of electroweak symmetry breaking. The production of high-energy muon beams at the intensities needed for these applications will require muon-beam cooling~\cite{Status-Report,Overarching-report,instr99,FFAG}.

\section{Cooling}
To accelerate a secondary or tertiary beam  
it is desirable first to decrease its size so that a reasonable fraction of the produced particles will fit inside the apertures of the beamline.
It is well known that a focusing element (\eg\ a pair of quadrupole magnets with opposed field gradients) can decrease the area of a charged-particle beam while increasing its spread in transverse momentum and, consequently, its divergence. This relationship is an example of Liouville's theorem: conservative forces cannot increase or decrease the volume occupied by a beam in six-dimensional phase space~\cite{nonlinear}.  

Focusing alone does not suffice for efficient acceleration of a secondary or tertiary beam, since the resulting increase in divergence means the beam will exceed some other aperture further downstream.  What is needed instead is a process by which {\em both} the beam size and divergence can be reduced. By analogy with refrigeration, which decreases the random relative motions of the molecules of a gas, this is known as beam {\em cooling}.

It is convenient to represent the volume of phase space occupied by a beam by the beam's {\em emittance}. The emittance in a given coordinate can be expressed as 
\begin{equation}
\epsilon_{i,n}\equiv \sigma_i\sigma_{p_i}/mc,
\label{eq:eps-n}
\end{equation}
where $\sigma$ designates root-mean-square, $i=x,y,z$,  and the factor $1/mc$ is introduced so as to express emittance in units of length ($m$ is the mass of the beam particle and $c$ the speed of light).  Neglecting possible correlations among the coordinates and momenta~\cite{covar}, we then have
$\epsilon_{6,n}=\epsilon_{x,n}\epsilon_{y,n}\epsilon_{z,n}$ for the six-dimensional emittance. The subscript $n$ distinguishes these {\em normalized} emittances from the frequently used {\em unnormalized} emittance
\begin{equation}
\epsilon_i\equiv\epsilon_{i,n}/\gamma\beta\,,
\end{equation} 
where $\gamma$ and $\beta$ are the usual relativistic factors. In terms of (unnormalized) emittance, the transverse beam sizes are given by 
\begin{equation}
\pi\sigma_x^2=\beta_x\epsilon_x\,,\qquad \pi\sigma_y^2=\beta_y\epsilon_y\,,
\end{equation}
where $\beta_x,\beta_y$ are the transverse amplitude functions of the focusing lattice in the $x$ and $y$ directions, which characterize the focusing strength along the lattice (low $\beta_i$ corresponds to strong focusing in the $i$th direction).

Since Liouville's theorem tells us that normalized emittance is a constant of the motion, beam cooling requires a ``violation" of
Liouville's theorem. 
This is possible by means of dissipative forces such as
ionization energy loss~\cite{Lichtenberg}, as described in more detail below.

\section{Muon Cooling}

Cooling of the transverse phase-space coordinates of a muon beam can be accomplished by passing the beam through energy-absorbing material and accelerating structures, both embedded within a focusing magnetic lattice; this is known as ionization cooling~\cite{early-cooling, Lichtenberg}.
Other cooling techniques (electron, stochastic, and laser cooling) are far too slow to yield a significant degree of phase-space compression within the muon lifetime~\cite{optical}. 
Ionization of the absorbing material by the muons decreases the muon momentum while (to first order) not affecting the beam size; by Eq.~\ref{eq:eps-n}, this constitutes cooling. At the same time, multiple Coulomb scattering of the muons in the absorber increases the beam divergence, heating the beam. Differentiating Eq.~\ref{eq:eps-n} with respect to path length, we find that the rate of change
of normalized transverse emittance  within the absorber is
given approximately by~\cite{Neuffer2,Fernow}
\begin{equation}
\frac{d\epsilon_n}{ds}\ =\
-\frac{1}{\beta^2} \left\langle\frac{dE_{\mu}}{ds}\right\rangle\frac{\epsilon_n}{E_{\mu}}\ +
\ \frac{1}{\beta^3} \frac{\beta_\perp (0.014)^2}{2E_{\mu}m_{\mu}L_R}\,,
\label{eq:cool}
 \end{equation} 
where angle brackets denote mean value, muon energy $E_\mu$  is in GeV, $\beta_\perp$ is evaluated at the location of the absorber, and $L_R$ is the radiation length of the absorber medium~\cite{Kim-Wang}.  (This is the expression appropriate to the cylindrically-symmetric case  of solenoidal focusing, where $\beta_x=\beta_y\equiv\beta_\perp$.) The first term in Eq.~\ref{eq:cool} is the cooling term and the second is the heating term.

To minimize the heating term, which is proportional to the $\beta$ function and inversely proportional to radiation length, it has been proposed~\cite{Status-Report} to use hydrogen 
as the energy-absorbing medium, giving $\langle dE_{\mu}/ds\rangle\approx 30\,$MeV/m and $L_R=8.7\,$m~\cite{PDG}, with superconducting-solenoid focusing to give small $\beta\sim10\,$cm~\cite{Lilens}. Key issues in absorber R\&D include coping with the large heat deposition by the intense ($\sim 10^{14}/$s)~\cite{FS2} muon beam and minimizing scattering in absorber-vessel windows~\cite{Abs-RD, Kaplan-NuFACT01,Kaplan-PAC2001,Kaplan-NuFACT00}.

\begin{figure}
\vspace{-.1in}
\centerline{\scalebox{0.9}{\includegraphics{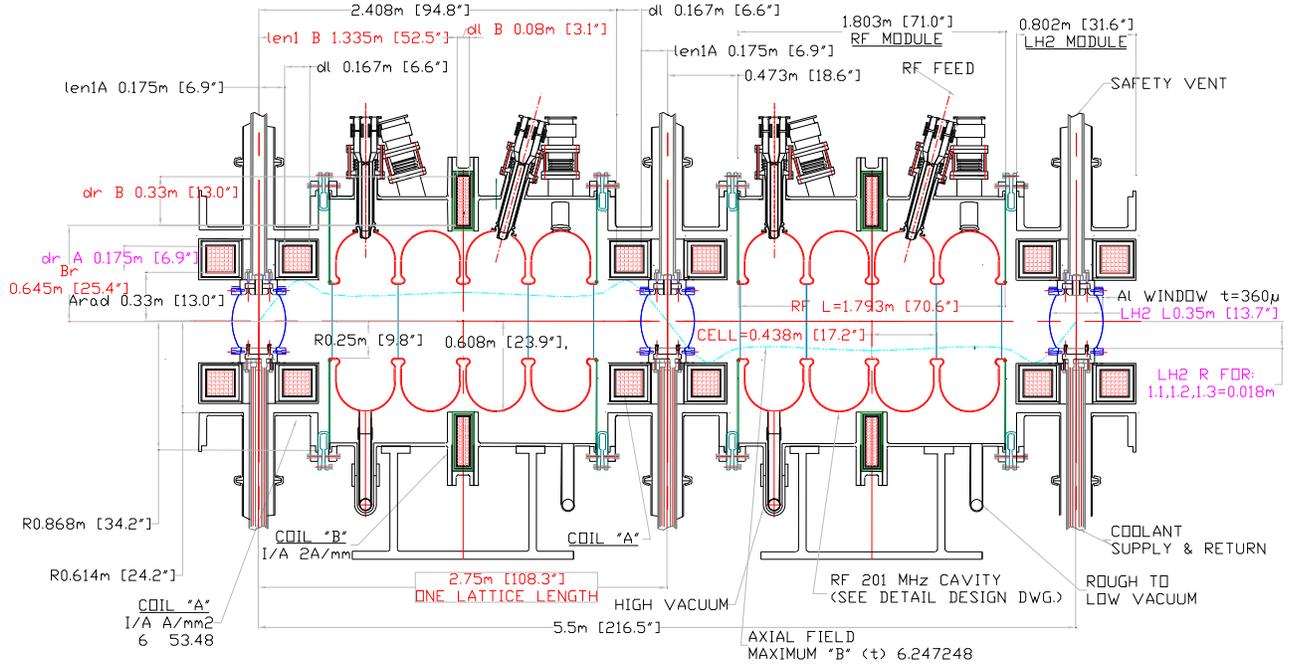}}}
\vspace{-.1in}
\caption{SFOFO ionization-cooling lattice (from \protect\cite{FS2}).}
\label{fig:SFOFO}
\end{figure}

An additional technical requirement is high-gradient reacceleration of the muons between absorbers to replace the lost energy, so that the ionization-cooling process can be repeated many times. Even though it is the absorbers that actually cool the beam, for typical radio-frequency (RF) accelerating-cavity gradients ($\sim 1-10\,$MeV/m), it is the RF cavities that dominate the length of the cooling channel (see \eg\ Fig.~\ref{fig:SFOFO}), and the achievable RF gradient determines how much cooling is practical before an appreciable fraction of the muons have decayed. We see from Eq.~\ref{eq:cool} that the percentage decrease in normalized emittance is proportional to the percentage energy loss, thus cooling in one transverse dimension by a factor 1/$e$ requires $\approx$100\% energy loss and replacement. The expense of RF power favors low beam energy for cost-effective reacceleration, as do also the increase of $\langle dE/dx\rangle$ and the decreased width of the $dE/dx$ distribution at low momentum~\cite{PDG}, and most muon-cooling simulations to date have used $\langle p \rangle\approx 200 - 300\,$MeV/$c$. We are pursuing R\&D on high-gradient normal-conducting RF cavities suitable for insertion into a solenoidal-focusing lattice~\cite{RF-RD}.

Transverse ionization cooling causes the longitudinal emittance $\epsilon_z$ to grow. Several effects contribute to this growth: fluctuations in energy loss in the absorbers (energy-loss straggling, or the ``Landau tail") cause growth in the energy spread of the beam, as does the negative slope of the $\langle dE/dx\rangle$ momentum dependence in the beam-momentum regime (below the ionization minimum) that we are considering~\cite{PDG}. Moreover, these low-momentum, large-divergence beams have a considerable spread in propagation velocity through the cooling lattice, causing bunch lengthening. These effects result in gradual loss of particles out of the RF bucket. They could be alleviated by longitudinal cooling.

Longitudinal ionization cooling is possible in principle, but it appears to be impractical~\cite{Snowmass96}. Its realization would call for operation {\em above} the ionization minimum, where the $\langle dE/dx\rangle$ slope with momentum is positive~\cite{PDG}, but that slope is small and the resulting weak cooling effect is overcome by energy-loss straggling. Instead what is envisioned is emittance {\em exchange} between the longitudinal and transverse degrees of freedom, decreasing the longitudinal emittance while at the same time increasing the transverse.  Conceptually, such emittance exchange can be accomplished by placing a suitably shaped absorber in a lattice location where there is dispersion, \ie, using a bending magnetic field to spread the muons out in space according to their momenta, and shaping the absorber so as to absorb more energy from the higher-momentum muons and less from the lower-momentum ones.  (One can see that this is emittance {\em exchange} rather than longitudinal cooling {\em per se}, since to the extent that the muon momentum spread has been reduced by the shaped absorber, the beam can no longer be reconverged to a small spot by a subsequent bend.) This is followed by transverse ionization cooling, the combined process being effectively equivalent to longitudinal cooling~\cite{emitt-exch}. 

\section{Cooling-channel designs}

A variety of focusing-lattice designs for transverse muon cooling have been studied, most using solenoids as focusing elements. Especially for the large ($\approx0.6\,$m) aperture required at the beginning of a muon cooling channel, stronger focusing gradients are possible using solenoids than using quadrupoles, and unlike quadrupoles, solenoids have the virtue of focusing both transverse dimensions simultaneously, giving a more compact lattice. 

While a high-field solenoid can produce a small (and constant) $\beta_\perp$, it is straightforward to see that a single such solenoid is not sufficient for muon cooling~\cite{Fernowetal}. A charged particle entering a solenoid off-axis receives a transverse magnetic kick from 
the fringe field, such that the particle's straight-line motion in the field-free region becomes helical motion within the solenoid. The exit fringe field must thus impart an equal and opposite kick so that the particle resumes its straight-line motion in the subsequent field-free region.

If within the solenoid the particle loses energy in an absorbing medium, the angular momentum of its helical motion must decrease, resulting in an imbalance between the entrance and exit kicks. The particle then exits the magnet with a net angular momentum, implying that a parallel beam entering an absorber-filled solenoid will diverge upon exiting. To cancel this net angular momentum, the field direction must alternate periodically.
The simplest case conceptually is focusing by a constant solenoidal field, but with one ``field flip" halfway along the cooling channel~\cite{Monroe}.  The length of a uniform section can be of order $10-100\,$m. Better performance can be achieved by adding a second field flip~\cite{double-flip}.
At an opposite extreme, the solenoidal-field direction can be flipped every meter or so, leading to a variety of solenoidal-focusing lattices dubbed alternating solenoid~\cite{Fernowetal,Status-Report,instr99,Kim-Yoon}, FOFO~\cite{Penn,Snowmass96,Monroe}, DFOFO~\cite{Kim-Yoon}, SFOFO (see Fig.~\ref{fig:SFOFO}), etc.~\cite{Penn,Wang}.

\section{Conclusions}
Detailed six-dimensional simulations show that enough transverse cooling can be achieved to build a
high-performance neutrino factory~\cite{FS2}, using either a double-flip or SFOFO cooling lattice.  For example, in Palmer's recent SFOFO design an initial transverse normalized emittance of $17\pi\,$mm$\cdot$rad is cooled in a 400-m-long cooling system to a final emittance of $2.8\pi\,$mm$\cdot$rad with $\approx$75\% muon loss~\cite{FS2,minicooling}. Such a facility would produce $\sim10^{20}$
neutrinos per year aimed at a far detector that could be thousands of km from the source, giving oscillation sensitivity at least two orders of magnitude beyond that of long-baseline experiments now under construction~\cite{FS2,Overarching-report}. Without longitudinal-transverse emittance exchange, transverse cooling reaches a point of diminishing returns as emittance growth in the longitudinal phase plane causes muons to be lost from the RF bucket. While emittance exchange would be helpful but not essential for a neutrino factory, to achieve the considerably smaller emittances required in a muon collider, it is mandatory.  R\&D on emittance exchange is ongoing, and several promising ideas are being actively explored~\cite{emitt-exch,workshops}.

\begin{acknowledgments}
This work was
supported in part by the U.S. Dept.\ of Energy, the National Science
Foundation, the Illinois Board of Higher Education, and the Illinois Dept.\ of Commerce and Community Affairs.
\end{acknowledgments}

\begin{thebibliography}{99}
\bibitem{covar}
More generally, letting  V be the covariance matrix of $(x,p_x,y,p_y,z,p_z)$, the six-dimensional normalized emittance is given by $\epsilon_{6,n}\equiv\det{\rm V}/(mc)^6$.
\bibitem{Lilens}
Another possibility is liquid-lithium lenses, which could serve as both the focusing element and the absorbing material. Prototype studies in Novosibirsk have so far 
been unsuccessful. A third possibility is high-pressure gaseous absorbers~\protect\cite{Kaplan-NuFACT01}.
\bibitem{minicooling}
The numbers quoted above include the ``minicooling" section, located upstream of the ``buncher," which precedes the SFOFO channel. In the $\approx$110-m-long SFOFO channel itself the emittance is cooled from $\approx$13$\pi$ to 2.8$\pi$\,mm$\cdot$rad  with $\approx$50\% muon loss.
\end{thebibliography}
\end{document}